\documentclass[aps,prb,twocolumn,showpacs,superscriptaddress]{revtex4} 
\usepackage{epsfig,amsmath,amssymb}
\usepackage{graphicx}
\usepackage[english]{babel}

\begin{document}

\title{The two Josephson junction flux qubit with large tunneling amplitude} %
\author{V.I. Shnyrkov}
\email{shnyrkov@ilt.kharkov.ua} %
\affiliation{B. Verkin Institute for Low Temperature Physics and Engineering, %
National Academy of Sciences of Ukraine, 61103 Kharkov, Ukraine} %
\author{A.A. Soroka}
\affiliation{National Science Center "Kharkov Institute of Physics
and Technology", Akhiezer Institute for Theoretical Physics, 61108 Kharkov, Ukraine} %
\author{S.I. Melnyk}
\affiliation{M.K. Yankel Kharkov National University of Radio
Electronics, Ministry of Education of Ukraine, 61161 Kharkov, Ukraine} %

\begin{abstract}
In this paper we discuss solid-state nanoelectronic realizations of
Josephson flux qubits with large tunneling amplitude between the two
macroscopic states. The latter can be controlled via the height and   
form of the potential barrier, which is determined by quantum-state
engineering of the flux qubit circuit. The simplest circuit of the
flux qubit is a superconducting loop interrupted by a Josephson
nanoscale tunnel junction. The tunneling amplitude between two
macroscopically different states can be essentially increased, by
engineering of the qubit circuit, if tunnel junction is replaced by
a ScS contact. However, only Josephson tunnel junctions are
particularly suitable for large-scale integration circuits and
quantum detectors with preset-day technology. To overcome this
difficulty we consider here the flux qubit with high-level energy
separation between "ground"$\,$ and "excited"$\,$ states, which
consists of a superconducting loop with two low-capacitance
Josephson tunnel junctions in series. We demonstrate that for real
parameters of resonant superposition between the two macroscopic
states the tunneling amplitude can reach values greater than 1$\,$K.
Analytical results for the tunneling amplitude obtained within
semiclassical approximation by instanton technique show good
correlation with a numerical solution.
\end{abstract}

\pacs{03.75.Lm, 74.50.+r, 85.25.Cp}

\maketitle

\section{Introduction}
Since successful demonstration of Rabi oscillations and Landau-Zener
coherent effects \cite{NPT,Vion,Yu,Martinis,Sillanpaa}, the
superconducting qubits (quantum bits) based on mesoscopic Josephson
junctions became the subject of consideration as possible candidates
to be the basic elements of a quantum computer hardware
\cite{Wendin,Makhlin}, including detectors to measure the state of
an individual qubit
\cite{Zorin1,Zorin2,Shnyrkov1,Shnyrkov,Il'ichev}. The Josephson
junction (JJ) qubits have two energy scales which are the Josephson
coupling energy $E_J$ and the charging energy $E_C$ of the JJ, and
they are subdivided into flux qubits, charge qubits, as well as
charge-phase qubits. In principle, all circuits of a quantum
computer can be fabricated by modern techniques using these
superconducting qubits. However, it is but poor quality
\cite{Wendin,Mooij} of the experimentally tested elements that is
the limiting factor on the way of implementation of quantum
registers. For example, an important but still unsolved problem in
the physics of a qubit working in the charge regime with $E_C/E_J\gg
1$ is an essential decrease of high spectral density of the noise
associated with the motion of charge in traps. In its turn, the
phase qubit ($E_J/E_C\gg 1$), which utilizes the phase of the
superconducting order parameter as a dynamic variable, is much less
sensitive to the charge fluctuations but is subject to the influence
of the noise in critical current of JJ, spin fluctuations and
Nyquist noise currents generated by excess ambient temperature. The
tunnel splitting of the energy levels arising from the coherent
superposition of the macroscopic states is small usually, $\Delta
E_{01}$ $\sim$ 150 -- 250 mK. Taking into account the effective
noise temperature, which can reach $T_{eff}$ $\sim$ 50 -- 100 mK in
experimental studies of the qubit dynamics, leads to a dramatic fall
of the decoherence times $\tau_\varphi$ and relaxation times
$\tau_\varepsilon$ \cite{Grifoni,Tian,Smirnov}. This means that, in
order to enhance considerably the qubit quality \cite{Wendin} (the
number of one-bit operations during the coherence timespan), the
system with large ($\Delta E_{01}\gtrsim 1$\,K) tunnel splitting of
the energy levels should be created.

Undoubtedly, the problem of creation of a quantum register based on
Josephson qubits brings up many issues but presently the invention
of a high-quality qubit is the most important one among them. It is
easy to show that the rate of the energy exchange between two
macroscopic states in a flux qubit is bounded by the "cosine"$\,$
shape of the potential barrier and cannot be increased owing to
decreasing the barrier height since the latter determines the
characteristic rate of thermal decay of the current-flow states. A
similar limitation associated with the lowering of the effective
barrier height can appear also when highly increasing the
pre-exponential factor. It is absolutely obvious that the ideal case
for a flux qubit is when the tunnel barrier in the phase space looks
like $\Pi$-shaped function having sufficiently large height and
small action. It was this issue that motivated the authors of the
Ref. \onlinecite{Mooij} for analyzing the phase-slip qubit, whose
creation required developing a new non-Josephson technology. In this
paper we search for an improved barrier design for the JJ flux
qubit.

The recent Ref. \onlinecite{Shnyrkov} demonstrated how the level
splitting can be increased at low temperatures ($T\rightarrow 0$) by
an order of magnitude with the potential barrier height kept
unchanged by modifying the qubit's potential barrier shape due to
using the clean-limit ScS junction in the superconducting ring.
However, the fabrication difficulties of obtaining pure and
reproducible ScS junctions are the serious hindrance in the way of
designing large-scale integrated qubit circuits.

\begin{figure}[t!]
\centering %
\vspace{3mm}
\includegraphics[width = 0.93\columnwidth]{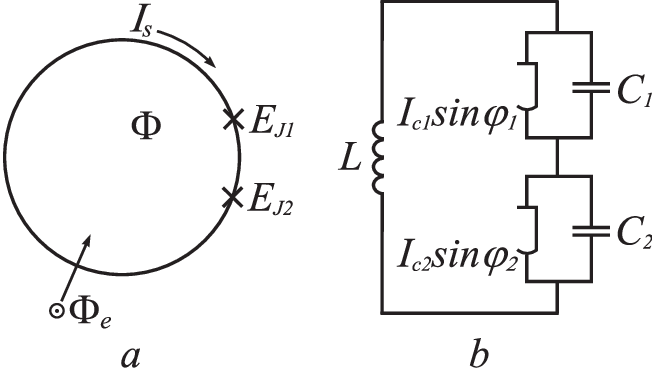} %
\caption{\label{fig01} Schematic picture of the proposed 2JJ flux
qubit with a SQUID configuration (a) and its circuit diagram (b).
The loop carrying supercurrent $I_s$ is pierced by an externally
applied magnetic flux $\Phi_e$ (towards an observer). The individual
SIS Josephson junctions are characterized by coupling energies
$E_{J1}, E_{J2}$, critical currents $I_{c1},I_{c2}$ and capacitances
$C_1, C_2$ which do not differ significantly. The loop inductance
$L$ is small enough so that the 2JJ SQUID has only
two metastable flux states.%
The parameter $g_0^{min}=E_{J}(\pi)/E_{C}\gg 1$ (see below). %
}
\end{figure}

To solve this problem, the analysis is carried out in the present
paper of the two-Josephson-junction flux qubit (2JJ flux qubit),
which can be considered as a superconducting ring of inductance $L$
interrupted by two almost equivalent tunnel SIS junctions with the
Josephson energies $E_{J1}, E_{J2}$, the critical currents
$I_{c1},I_{c2}$ and the capacitances $C_1, C_2$, respectively (see
Fig.\;\ref{fig01}(a,b)). The difference between the two SIS
mesoscopic junctions will be characterized by the asymmetry
parameter $\lambda = I_{c2}/I_{c1} = E_{J2}/E_{J1} = C_2/C_1\le 1$
so that the "junction 1"$\,$ would have greater or equal values of
the Josephson energy, the critical current and the capacitance as
compared to the "junction 2"$\,.$ The external magnetic flux
$\Phi_e$ can be coupled to the qubit by a separate coil located in
close proximity to the qubit's loop. It is well known that in
classical limit the circulating current $I_s$ as a function of
external magnetic flux for dc SQUIDs with $I_{c1}=I_{c2}$ has the
singularity in the points of $\Phi_e=\Phi_0(n+1/2)$ ($\Phi_0$ is the
flux quantum) so that the two-junction interferometer can be
considered as a "single-junction"$\,$ one, with the potential energy
shape in the phase space being modified. Below we indicate
conditions for the proposed 2JJ flux qubit under which the classical
Josephson relationship between phase differences on JJ contacts is
retained in the quantum regime, phase (flux) is a good quantum
variable and the charging effect on the island between JJ contacts
is negligible.

The problem lies in determining and analyzing the tunnel splitting
$\Delta E_{01}=E_1-E_0$ of the degenerate zero energy level in the
double-well symmetric potential of a 2JJ flux qubit (at
corresponding external conditions) resulted from the coherent
quantum tunneling of the magnetic flux between the wells. In the
proposed mesoscopic system in the quantum regime, the two lower
energy levels $E_0$ and $E_1$ arising from coherent superposition of
the macroscopically distinct flux or persistent-current states form
a qubit. It turns out that, because of the change in the form of the
potential energy of the 2JJ flux qubit as compared to the 1JJ qubit,
the tunnel splitting $\Delta E_{01}$ can multiply rise reaching the
values $\gtrsim1\,$K (in temperature units) and substantially
enhance the properties of the qubit as a basic element
for quantum computations. %
The sensitivity of the $\Delta E_{01}$ magnitude to $\lambda$ as
well as to the junction parameters can limit applications based on
the 2JJ flux qubit both for quantum computation and quantum
detectors.

\section{Theoretical model and results}
We will discuss the 2JJ flux qubit in the approximation of the
Hamiltonian of an isolated system in the zero temperature limit. All
the dissipative processes associated with the own and the external,
regarding the system, degrees of freedom (the quasiparticles, the
magnetic flux fluctuations in the qubit and in the outer measuring
circuit, etc.) are neglected in this approximation. In the framework
of this approximation, only the supercurrent component flows in the
qubit ring which in the classical regime, according to the Josephson
relation, is equal to
\begin{equation}\label{I_s}
I_s = I_{c1}\sin\varphi_1=I_{c2}\sin\varphi_2,
\end{equation}
where $\varphi_1, \varphi_2$ are the order parameter phase
differences at corresponding tunnel junctions. It is convenient to
count the values of the supercurrent $I_s$ and the phase differences
at the junctions clockwise, the applied magnetic flux $\Phi_e$, the
total magnetic flux in the ring $\Phi$ and the supercurrent $I_s$
being tied by the relation $\Phi=\Phi_e-LI_s(\Phi)\,.$ The classic
Hamiltonian of the 2JJ flux qubit in the approximation of the
isolated system contains the contributions of the electrostatic
energy of the charges in the junction capacitances, the junction
Josephson energies and the magnetic energy of the supercurrent in
the ring, and has the form:
\begin{equation}\label{H-phys}
\begin{array}{l}
\displaystyle{ H=\frac{(2eN_0)^2}{2}\left( \frac{1}{C_1} + \frac{1}{C_2} \right) - }\\ %
\displaystyle{-\left(E_{J1}\cos\varphi_1 + E_{J2}\cos\varphi_2\right) 
+ \frac{(\Phi - \Phi_e)^2}{2L}+E_{0}\,, }
\end{array} 
\end{equation}
where $N_0$ is the number of the excess (deficient) Cooper pairs in
the banks of the SIS Josephson junctions, $E_{0}$ is the constant
fixing the reference level for the potential energy. Using relation
(\ref{I_s}), we will reduce expression for the Josephson energy in
classic Hamiltonian (\ref{H-phys}) to the form
$U_J^{0}(\phi)=-E_J(\phi) = -(E_{J1}\cos\varphi_1 +
E_{J2}\cos\varphi_2) = \displaystyle{- E_{J1}\sqrt{(1-\lambda)^2 +
4\lambda\cos^2(\phi/2) } }\,,$ where a new variable of the overall
phase $\phi = \varphi_1 + \varphi_2$ is introduced.

The proposed 2JJ qubit system is topologically analogous to the
charge-phase qubit \cite{Zorin1,Krech}, representing a
single-Cooper-pair tunneling transistor (SCPT-transistor consists of
two Josephson junction contacts with the voltage gate next to the
island between them) inserted in a superconducting ring. Therefore
the structure of the Josephson energy in the Hamiltonian
(\ref{H-phys}) of the 2JJ qubit is similar to that of the
charge-phase qubit. The main difference, affecting the Josephson
energy form, lies in that: (i) in the 2JJ qubit there is no charge
gate and no polarization charge $Q_0$ is induced through it on the
island; (ii) the charge-phase qubit is designed to work in the
charge mode, whereas the 2JJ qubit is designed to work in the flux
mode, that is in an opposite extreme dynamic regime. %
In the preceding Ref.\;\onlinecite{Zorin2}, devoted to a quantum
detector based on the SCPT-transistor, its working regimes were
investigated that depend on the form of the
Josephson energy of the system (formulae (1),(2) in Ref.\;\onlinecite{Zorin2}: %
$U_J(\phi,\varphi)= -(E_{J1}\cos\varphi_1 +
E_{J2}\cos\varphi_2)=-E_J(\phi)\cos(\varphi+\gamma(\phi))=-E_J(\phi)\cos\chi,$
$\varphi=(\varphi_1-\varphi_2)/2,
\tan\gamma(\phi)=\frac{\lambda-1}{\lambda+1}\tan(\phi/2)$), and can
be characterized by the parameter $g_0=E_{J}(\phi)/E_{C}$, where
$E_{C}=e^2/2C$ is the characteristic charging energy of the island
between JJ contacts, $C= C_1+C_2+C_g$ being the total capacitance of
the island regarding the rest of the system ($C=C_1+C_2$ for the 2JJ
qubit as $C_g=0$). The parameter $g_0$ crucially determines the
mutually conditioned quantum-averaged supercurrent (current-phase
dependence) $I_s(\phi)$ and effective Josephson energy $U_J(\phi)$
of the SCPT-transistor and based on it charge-phase qubit
respectively \cite{Zorin1}. At solving the Schr\"odinger equation,
the supercurrent $I_s(\phi)=I_s^0(\phi)\langle\cos\chi\rangle$ is
represented as appropriate supercurrent in the classical limit
$I_s^0(\phi)$, multiplied by the function $\langle\cos\chi\rangle$
that describes an effective influence of charge fluctuations on the
island between JJ contacts (formulae (4),(5) and Fig. 2(a) with a
family of dependencies $\langle\cos\chi\rangle(Q_0)$ on the
parameter $g_0$ in Ref.\;\onlinecite{Zorin2}). This result of
Ref.\,\onlinecite{Zorin2} reflects a physically clear conclusion:
(i) the effect of fluctuations of Cooper-pair number
($\hat{n}=-id/d\varphi$) on the island that affects $I_s(\phi)$ and
$U_J(\phi)$ is well apparent in the charge mode of system dynamics
($g_0\lesssim 1$), at that the function
$\langle\cos\chi\rangle(Q_0)$ being strongly reduced and modulated;
(ii) in the opposite limit $g_0\gg 1$ the function
$\langle\cos\chi\rangle$ becomes a constant close to unity, and at
$g_0\rightarrow \infty$, $\langle\cos\chi\rangle\equiv 1$ so that in
this limit the current-phase dependence $I_s(\phi)=I_s^0(\phi)$ and
the effective Josephson energy $U_J(\phi)=U_J^{0}(\phi)=-E_J(\phi)$
are described by classical expressions. It is interesting to note
that experimental investigation of the charge-phase qubit with the
parameter $g_0\sim 1$ ($E_J(\phi)\sim E_C$, i.e. $E_J(\phi)$ rather
large for a charge-phase qubit) at low temperatures (20\,mK)
demonstrated \cite{Shnyrkov_exp} the influence of gate quasicharge
$Q_0$ of the order of noise, and characteristic form of the
current-phase dependence qualitatively conforming to an appropriate
classical dependence (see Eq.\,(\ref{I-phi}) below).

We consider the extreme case $g_0\gg 1$ to realize the 2JJ qubit to
work in the flux mode, where the effect of charge fluctuations due
to the variable $\varphi$ is negligible (so that $\varphi$ falls out
from the Hamiltonian). In this extreme case the classical
expressions for the Josephson energy of the two-junction
interferometer and for the supercurrent through its loop apply to
the quantum regime of system dynamics, and the relationship
(\ref{I_s}) between variables $\varphi_1,\varphi_2$ holds.

Note that for 1JJ flux qubit the usual condition of phase being a
good quantum variable is defined by the parameter
$g=E_{J1}/E_{C1}\gg 1$, where $E_{C1}=e^2/2C_1$ is characteristic
electrostatic energy of the JJ contact. Then the minimum value of
the parameter $g_0$ of the 2JJ flux qubit as a system and the
parameter $g$ being characteristic of a single JJ contact of the
qubit are connected by the relation
$g_0^{min}=E_{J}(\pi)/E_{C}=E_{J1}(1-\lambda)/[e^2/2C_1(1+\lambda)]=(1-\lambda^2)
g$. Thus, the 2JJ flux qubit have to satisfy the condition
$g_0^{min}=(1-\lambda^2) g\gg 1,$ and the parameter $\lambda$ is
bounded from above by this condition.

Due to the single-valuedness of the superconducting order parameter
the variable $\phi$ satisfies the condition
\begin{equation}\label{Fi}
\phi= \varphi_1 + \varphi_2 = 2\pi\frac{\Phi}{\Phi_0}+2\pi n,\,\Phi
= n\Phi_0 + \frac{\phi}{2\pi}\Phi_0,\, \Phi_0=\pi\hbar/e\,,
\end{equation}
where $n$ is the integer number of the flux quanta $\Phi_0$ in the
total magnetic flux $\Phi$ (below, we will consider the qubit to
work in the $n=0$ mode). Owing to the relationships (\ref{I_s}),
(\ref{Fi}) there is the only independent phase variable from
$\varphi_1,\varphi_2,\phi,$ and in the quantum regime the physical
fluctuating quantum variable is the total phase difference $\phi$ at
the both junctions which equals, to within $2\pi$ factor, to the
total magnetic flux in the ring in the units of flux quantum
($\phi/2\pi=\Phi/\Phi_0$). %

The transition to the quantum description of the flux qubit consists
in associating the value $N_0$ of the Cooper pairs tunneling through
the junctions with the operator $\displaystyle{\hat{N}_0 =
-i\frac{\partial}{\partial\phi}\,,}$ conjugated to the phase
operator $\hat{\phi}$ ($[\hat{N}_0,\hat{\phi}]=-i$), and solving the
Schr\"odinger equation with the obtained quantum Hamiltonian in
$\phi$-representation \cite{Leggett}. By applying the quantization
procedure to the Hamiltonian (\ref{H-phys}) and writing down the
energy contributions via the variable $\phi$, we will come to a
canonical form of the Hamiltonian of the 2JJ flux qubit in the
quantum case:
\begin{equation}\label{H-quantum}
\begin{array}{l}
\displaystyle{ %
\hat{H}=\frac{\hat{P}^2}{2M} + \hat{U}(\phi)=
-\frac{\hbar^2}{2M}\frac{\partial^2}{\partial\phi^2} + }\vspace{0.15cm} \\%
\displaystyle{ %
+E_{J1}\left( \varepsilon_0 - \sqrt{(1-\lambda)^2 + 4\lambda\cos^2\frac{\phi}{2} } + %
\frac{(\phi-\phi_e)^2}{2\beta_L}\right)\,,}
\end{array}
\end{equation}
which can be considered as the Hamiltonian of a quantum particle
with the mass $M$ moving in the potential $U(\phi).$ Here
$\displaystyle{\hat{P}=\hbar
\hat{N}_0=-i\hbar\frac{\partial}{\partial\phi} }$
$([\hat{P},\hat{\phi}]=-i\hbar)$ corresponds to the particle
momentum operator,
$\displaystyle{M=\left(\frac{\Phi_0}{2\pi}\right)^2\frac{\lambda
C_1}{(\lambda+1)}}$ is its mass,
$\beta_L=\displaystyle{\frac{2\pi}{\Phi_0}LI_{c1}}=
\left(\frac{2\pi}{\Phi_0}\right)^2 L E_{J1}$ is the potential
parameter, $\displaystyle{\phi_e=2\pi\frac{\Phi_e}{\Phi_0}}$ is the
external magnetic flux parameter (the constant $\varepsilon_0$ is
chosen further from the condition for the symmetric potential of
equalling to zero in the minima points). The 2JJ flux qubit
parameter $g_0^{min}=(1-\lambda^2)\displaystyle{\frac{\beta_L}{L}
\left(\frac{\Phi_0}{2\pi}\right)^2\frac{2C_1}{e^2}=}
(1-\lambda^2)\beta_L\frac{C_1}{L}\frac{\hbar^2}{2e^4}\gg 1$.

\begin{figure}[t!]
\centering %
\vspace{3mm}
\includegraphics[width = 0.90\columnwidth]{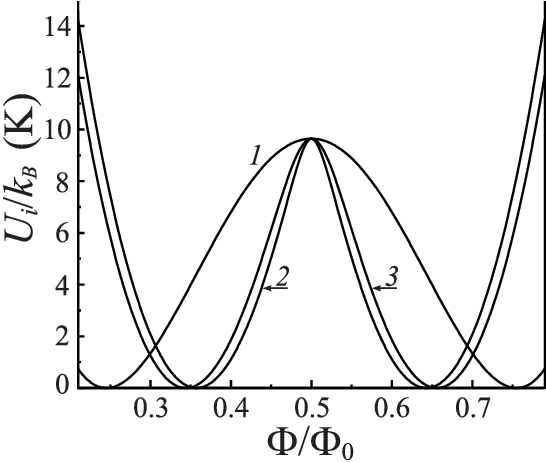} %
\caption{\label{fig02} Potential $U(\Phi/\Phi_0)/k_B$ in temperature
units for 1JJ qubit with $\beta_L=1.602$  -- {\it 1} and for 2JJ
qubit for such parameter couples ($\lambda,\beta_L$): $(0.9,1.058)$
-- {\it 2}, $(0.8,1.276)$ -- {\it 3} at external magnetic flux
$\Phi_e=\Phi_0/2.$ The geometric ring inductance is
$L=3.0\cdot10^{-10}$ H for both qubits; the potential barrier
heights $U_0$ in curves {\it 1} -- {\it 3} are equal,
$U_0/k_B=9.64\,$K.  }
\end{figure}

The potential $U(\phi)$ shape depends on the parameters $\lambda,
\beta_L, \phi_e.$ We are interested in the case of symmetric
potential which, according to (\ref{H-quantum}), is realized at
$\phi_e=\pi$ ($\Phi_e=\Phi_0/2$). It should be noticed that formally
in the extreme case of identical junctions, at $\lambda=1$ (though
really the value of $\lambda \approx 1$ should be such that to
satisfy the condition $g_0^{min}\gg 1$), the potential $U(\phi)$
coincides with the potential of the flux qubit based on the clean
ScS contact studied in Ref. \onlinecite{Shnyrkov} (Equation (3)):
$$U_{ScS}(\phi)=E_{J}\left( -2\left|\cos(\phi/2)\right| + %
\frac{(\phi-\phi_e)^2}{2\beta_L}\right)\,.$$ %
At the same time, owing to renormalization of the mass $M$ for the
2JJ flux qubit by the factor $\lambda/(\lambda+1)$ in respect to the
corresponding mass for the ScS flux qubit (provided that the
capacitances of SIS and ScS junctions are equal, $C_1=C$), for
$\lambda\approx 1$, the relation for masses is $M_{2JJ}\approx
M_{ScS}/2\,.$ Hence, at $\lambda\approx 1$ the splitting $\Delta
E_{01}$ in 2JJ qubit is expected yet more than in the ScS qubit. The
parameter $\beta_L$ determines the height of the potential barrier
of the double-well potential, so that the barrier height goes down
while reducing $\beta_L$. Like in the case of ScS qubit, the 2JJ
qubit potential has two local minima even at $\beta<1$ (unlike the
SIS qubit where the double-well potential exists at $\beta>1$ only),
which gives a possibility of considerable scaling down the geometric
dimension (inductance) of the system with the mesoscopic junctions.

Fig.\;\ref{fig02} shows the potential $U(\Phi/\Phi_0)/k_B$ of 2JJ
flux qubit for two parameters couples ($\lambda,\beta_L$) and also,
for the comparison sake, the well-known potential
$U_{SIS}(\Phi/\Phi_0)/k_B$ of 1JJ flux qubit at external magnetic
flux $\Phi_e=\Phi_0/2.$ The inductances $L$ for both types of the
qubits can be supposed equal (to specify the magnetic flux
fluctuation level) while the parameter $\beta_L$ (i.e. the critical
currents $I_{c1},I_c$ of the corresponding SIS junctions) in all the
dependences is chosen in such a way so that the potential barriers
in all the potentials were of the same height $U_0$. The latter
requirement implies roughly equal decay rates for the metastable
states due to thermal fluctuations, taking them into account being
beyond our consideration. Apparently, to realize the quantum regime
in a physical experiment, the value $U_0/k_B$ must highly exceed the
system temperature. As seen from Fig.\;\ref{fig02}, the potentials
$U(\Phi/\Phi_0)/k_B$ for a 2JJ qubit have lesser width (between the
potential minima points) as compared to %
the corresponding potentials for a 1JJ qubit while the area under
the potential curve between the points of its minima for the 2JJ
qubit shrinks greatly against the corresponding area for the 1JJ
qubit. Additionally, if the corresponding capacitances of the SIS
junctions in both 2JJ and 1JJ qubits are equal ($C_1=C$) then the
ratio of the effective masses for these qubits is
$\lambda/(\lambda+1)$. As it will be shown below, it is the change
in the potential shape and the decrease of the effective mass in 2JJ
qubit that lead to multiple rise in the amplitude of its tunnel
splitting. %

\begin{figure}[t!]
\centering %
\vspace{3mm}
\includegraphics[width = 0.90\columnwidth]{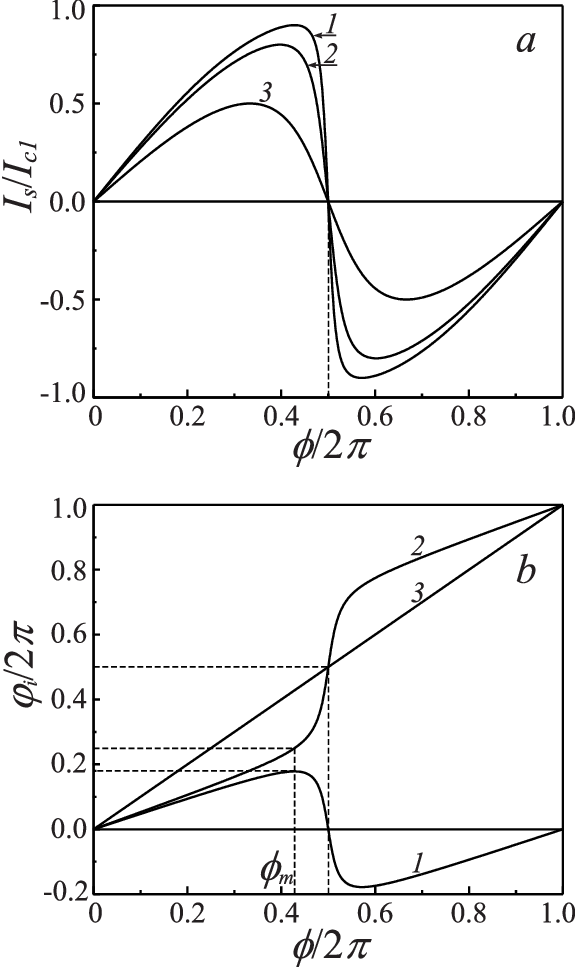} %
\caption{\label{fig03} a) Integral phase-current relation
$I_s(\phi/2\pi)/I_{c1}$ for 2JJ qubit at various $\lambda$: $0.9$ --
{\it 1}, $0.8$ -- {\it 2}, $0.5$ -- {\it 3}.\; %
b) Functions $(\varphi_1/2\pi)[\phi/2\pi]$ -- {\it 1},
$(\varphi_2/2\pi)[\phi/2\pi]$ -- {\it 2} for 2JJ qubit at
$\lambda=0.9$. The straight line
$\varphi_1(\phi)+\varphi_2(\phi)=\phi$ -- 3 corresponds to the
$\phi$ definition. The values of
$\varphi_1/2\pi=\arcsin(\lambda)/2\pi\approx 0.18$ and
$\varphi_2/2\pi=0.25$ (the latter being $\lambda$-independent)
correspond to $\phi_m/2\pi=\arccos(-0.9)/2\pi\approx 0.43$. }
\end{figure}

The current-phase relation for 2JJ qubit, directly related to the
Josephson potential energy $U_J(\phi)$, is derived from (\ref{I_s}),
(\ref{Fi}):
\begin{equation}\label{I-phi}
\displaystyle{  %
\frac{I_s(\phi)}{I_{c1}} = \sin\varphi_1 = \lambda\sin\varphi_2 =
\frac{\lambda\sin\phi}{\sqrt{(1-\lambda)^2 + 4\lambda\cos^2(\phi/2)
} }\,. }
\end{equation}
The current-phase relation $I_s(\phi)$ extrema (which are equal by
their absolute values) are located in the points $\phi_m =
\arccos(-\lambda)$ (maximum; $\pi/2 \le \phi_m \le \pi$) and
$\phi_{m1} = 2\pi -\arccos(-\lambda)$ (minimum) symmetrically around
the point $\phi=\pi,$ where the supercurrent vanishes to zero
($I_s=0$) alternating its direction. Thus, at $\lambda$ being near
the unity, in the interval $(\phi_m,\phi_{m1})$, the supercurrent
$I_s$ changes dramatically from its maximum to minimum value with
alternating the current direction in the point $\phi=\pi.$
Fig.\;\ref{fig03}(a) displays the integral current-phase dependence
$I_s(\phi/2\pi)/I_{c1}$ for 2JJ qubit for several parameters
$\lambda$. The interval $(\phi_m,\phi_{m1})$ shrinks as the
parameter $\lambda$ increases and the maximum-to-minimum by-current
transition becomes more sharp (the extreme case $\lambda=1$ being
valid in classical SQUID dynamics corresponds to
$\phi_m=\phi_{m1}=\pi$ with the infinite derivative of the
current-phase relation in the point $\pi$). Let us also consider the
order parameter phase differences $\varphi_1(\phi),$
$\varphi_2(\phi),$ derived directly from (\ref{I-phi}). The analysis
of formula (\ref{I-phi}) shows that the function $\varphi_1(\phi)$
for a junction with high critical current has extrema in the points
$\phi_m,\phi_{m1}$. The transition from the maximum positive value
$\varphi_1(\phi_m)=\arcsin \lambda$ ($0\le \varphi_1(\phi_m)\le\pi$)
to the minimum negative value $\varphi_1(\phi_{m1})=-\arcsin
\lambda$ with alternating the phase difference sign in the point
$\pi$ ($\varphi_1(\pi)=0$) takes place in the interval
$(\phi_m,\phi_{m1})$, and $\varphi_1(0)=\varphi_1(2\pi)=0$. %
The function $\varphi_2(\phi)$ for a junction with lower critical
current is a monotonically increasing one from $\varphi_2(0)=0$ to
$\varphi_2(2\pi)=2\pi,$ which is symmetrical with respect to the
line $y=\phi$; $\varphi_2(\pi)=\pi,$ and $\varphi_2(\phi_m)\equiv
\pi/2,$ $\varphi_2(\phi_{m1})\equiv 3\pi/2.$ %
For the classical 2JJ SQUID, in the extreme case of $\lambda=1$ the
functions $\varphi_1(\phi),$ $\varphi_2(\phi)$ behave as follows:
$\varphi_1=\varphi_2=\phi/2$ at $0\le\phi<\pi$; at the point $\pi$ a
jump appears in the function $\varphi_1(\phi)$ between the values
$\pi/2,-\pi/2$ with further linear rise up to $\varphi_1(2\pi)=0,$
while the function $\varphi_2(\phi)$ demonstrates a jump between the
values $\pi/2,3\pi/2$ with further linear increase up to
$\varphi_2(2\pi)=2\pi.$ Fig.\;\ref{fig03}(b) exhibits dependences
$(\varphi_1/2\pi)[\phi/2\pi],$ $(\varphi_2/2\pi)[\phi/2\pi]$ for a
certain $\lambda,$ with their distinctive appearance. The straight %
line $\varphi_1(\phi)+\varphi_2(\phi)=\phi$ corresponds to the
$\phi$ definition showing the expansion of the total phase
difference over the both junctions into the component phase
differences of the order parameter over each of them.

We will find the tunnel splitting $\Delta E_{01}$ of the degenerate
zero level in the symmetrical (at $\phi_e=\pi$) double-well
potential $U(\phi)$ in the 2JJ flux qubit by numerical solution of
the Schr\"odinger equation and analytically by using instanton
technique in the semiclassical approximation. To find a numeric
solution of the Schr\"odinger stationary equation
\begin{equation}\label{Shred}
\centering %
\displaystyle{ \hat{H}\Psi(\phi)=E\Psi(\phi) }
\end{equation}
with Hamiltonian (\ref{H-quantum}), a kind of the finite elements
method is used with approximation of the potential $U(\phi)$ by a
piecewise constant function. Zero boundary conditions are used for
the wave function $\Psi(\phi)$, the domain width and the element
quantity being set so that provide good accuracy of the calculation.

In the semiclassical approximation the problem of a tunneling
quantum particle can be solved using the instanton technique
\cite{Coleman,Kleinert}. For a particle of the mass $M,$ moving at
zero temperature in symmetric double-well potential $V(x),$
referenced from its minimum level ($V(\pm a)=0,$ $\pm a$ are the
minimum points), the expressions for the energy levels $E_{1,0}$ and
the tunnel splitting $\Delta E_{01}$ read like
\begin{equation}\label{DE}
\begin{array}{l}
\displaystyle{ E_{1,0} = E_0 \pm \frac{\Delta E_{01}}{2} =
\frac{\hbar\omega_0}{2} \pm \hbar K \exp \left( -\frac{S_0}{\hbar}
\right)\,,}\vspace{0.1cm} \\
\displaystyle{S_0 = \int_{-a}^a dx \sqrt{2MV(x)}\,,}\vspace{0.1cm} \\
\displaystyle{K=A\omega_0 \sqrt{\frac{M\omega_0 a^2}{\pi\hbar}}\,,\;\;
\omega_0^2 = \frac{V''(\pm a)}{M}\,, }\vspace{0.1cm}\\  %
\displaystyle{ \Delta E_{01} = 2A\,\hbar \omega_0\, \sqrt{
\frac{M\omega_0 a^2}{\pi\hbar} } \exp \left( -\frac{S_0}{\hbar}
\right)\,. }
\end{array}
\end{equation}
Here $\omega_0$ is the frequency of the particle zero oscillations
in each of the wells, $S_0$ is the particle action on the instanton
trajectory, the dimensionless constant $A$ is found from the
equation for the instanton's function $t(x)$ in the asymptotic
limit:
\begin{equation}\label{A}
\begin{array}{l}
\displaystyle{ %
t(x)|_{x\rightarrow a}=\int_0^{x\rightarrow a}\frac{dx}{\sqrt{2V(x)/M} } =
-\frac{1}{\omega_0}\ln\frac{a-x}{Aa}\,, }\vspace{0.1cm} \\ %
\displaystyle{ %
A = \lim_{x\rightarrow a}\frac{(a-x)}{a}\exp\left( \int_0^{x} dx
\sqrt{\frac{M\omega_0^2}{2V(x)} }\right)\,. }
\end{array}
\end{equation}

Starting from the Hamiltonian (\ref{H-quantum}) and using formulae
(\ref{DE}),(\ref{A}), we obtain the tunnel splitting $\Delta E_{01}$
for the 2JJ flux qubit in the case of symmetric double-well
potential: %
\vspace{0.5cm} %
\widetext
\begin{subequations}\label{DE-2jj}
\begin{align}
& \displaystyle{ \Delta E_{01} = 4A\,\hbar \omega_0\,
\sqrt{ \frac{M\omega_0 \alpha_0^2}{\pi\hbar} } \exp \left( -\frac{S_0}{\hbar} \right)\,, }\\ %
& \displaystyle{ \frac{S_0}{\hbar} = \frac{2\hbar}{e^2}\sqrt{
\frac{\lambda
C_1}{(\lambda+1)L} } %
\int_0^{\alpha_0} d\alpha %
\sqrt{ \beta_L\left( \sqrt{ (1-\lambda)^2/4 + \lambda\sin^2\alpha_0}
- \sqrt{ (1-\lambda)^2/4 +  \lambda\sin^2\alpha} \right) + \alpha^2 - \alpha_0^2 }\,,  }\\ %
& \displaystyle{ \omega_0 = \sqrt{ \frac{(\lambda+1)}{\lambda C_1L}
\left( 1-\left.\frac{\beta_L}{2}\frac{d^2}{d\alpha^2}\sqrt{
(1-\lambda)^2/4 +  \lambda\sin^2\alpha}\right|_{\alpha_0}
\right)}\,,\quad M=\left(\frac{\Phi_0}{2\pi}\right)^2\frac{\lambda C_1}{(\lambda+1)} \,,}\\ %
& \displaystyle{ %
A = \lim_{\alpha\rightarrow
\alpha_0}\frac{(\alpha_0-\alpha)}{\alpha_0} \exp\left(
\sqrt{\frac{M\omega_0^2}{E_{J1}}} \int_0^{\alpha} %
\frac{d\alpha}{\sqrt{ \sqrt{  (1-\lambda)^2/4 +
\lambda\sin^2\alpha_0 } - \sqrt{ (1-\lambda)^2/4 +
\lambda\sin^2\alpha} + (\alpha^2 - \alpha_0^2)/\beta_L  }}
\right)\,, }\\
& 2\alpha_0\sqrt{ (1-\lambda)^2/4 +  \lambda\sin^2\alpha_0
}-\beta_L\lambda\sin\alpha_0\cos\alpha_0=0\,.
\end{align}
\end{subequations}
\endwidetext

\begin{figure}[t!]%
\centering %
\vspace{3mm}
\includegraphics[width = 0.90\columnwidth]{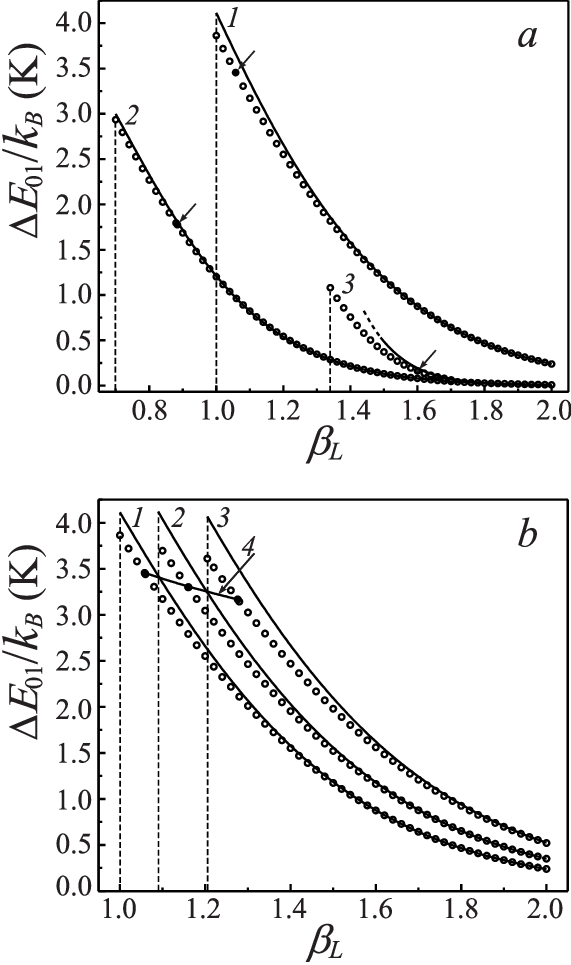} %
\caption{\label{fig04} a) Function $\Delta E_{01}(\beta_L)/k_B$ for
2JJ qubit at $\lambda=0.9$ -- {\it 1}, ScS qubit -- {\it 2}, 1JJ
qubit -- {\it 3}; the points in the numerical curves corresponding
to equal height (9.64 K) of the potential barrier for all the qubits
are indicated by arrows.\; %
b) Function $\Delta E_{01}(\beta_L)/k_B$ for 2JJ qubit and various
$\lambda$: $0.9$ -- {\it 1}, $0.85$ -- {\it 2}, $0.8$ -- {\it 3} and
"level line"$\,$ of equal heights (9.64 K) of the potential barriers
at varying $\lambda$ -- {\it 4}. The numerically obtained results
are pointed by hollow circles, the analytically obtained ones are
plotted by solid lines in the graphs (a) and (b). Dashed lines show
the lowest boundary $\beta_L$, at which the level height $E_1$
becomes equal to the potential barrier height $U_0$. %
For 1JJ and ScS qubits, the capacitance of corresponding (SIS and
ScS) junctions is $C=2.7\cdot10^{-15}$ F while for 2JJ qubit the
capacitance of the larger SIS junction is $C_1=2.7\cdot10^{-15}$ F.
The geometric ring inductance is $L=3.0\cdot10^{-10}$ H, the
parameter $g\approx 76\beta_L$ for all the qubits. For 2JJ flux
qubit the parameter $g_0^{min}\approx 76\beta_L(1-\lambda^2)\gtrsim
10$ at $\lambda \lesssim 0.93$ . }
\end{figure}

A variable $\alpha =(\phi-\pi)/2$ is introduced in formulae
(\ref{DE-2jj}) (due to the potential symmetry condition
$\phi_e=\pi$), the minima point of $\alpha_0>0$ of the potential
$U(\alpha)$ satisfying equation (\ref{DE-2jj}e). The accuracy of the
semiclassical approximation is high provided that $S_0/\hbar \gg 1,$
the method accuracy degrades as the dimensionless variable
$S_0/\hbar$ diminishes approaching the unity. The results of a
numerical analysis is of great importance in this region.

Fig.\;\ref{fig04}(a,b) presents the $\beta_L$-dependences of the
tunnel splitting $\Delta E_{01}(\beta_L)/k_B$ for 2JJ, ScS and SIS
flux qubits at the equal capacitances of the corresponding junctions
$C_1=C=2.7$ fF and at the inductance $L=0.3$ nH of the qubits loop.
In both plots the curves calculated numerically are pointed by
hollow circles while the ones obtained analytically using the
instanton technique are plotted by solid lines. The formulae
(\ref{DE-2jj}) were used for 2JJ qubit while similar formulae were
taken for ScS and SIS qubits based on the forms of their potentials.
The change in the parameter $\beta_L$ means the variation of the
critical currents $I_{c1},I_c$ of the corresponding junctions at a
fixed inductance $L$. The double-well potential height decreases 
with lowering the parameter $\beta_L$, the energy level $E_1$ being
equalized to the potential barrier height $U_0$ at a certain
$\beta_{L0}$ ($E_1=U_0$) and exceeding it with further $\beta_L$
lowering. Then, the wave function corresponding to the level $E_1$
is no further a superposition of the states localized in the left
and right wells. The boundary values $\beta_{L0}$ for the curves in
the figure are indicated by dash lines. In the vicinity of
$\beta_{L0},$ at $(U_0-E_1)\sim k_BT$, the quantum coherence will be
destroyed due to thermal fluctuation causing the over-barrier
transitions. One can see from Fig.\;\ref{fig04} that the numerically
and the analytically obtained curves almost coincide at large
$\beta_L$ and begin to diverge at lower $\beta_L$. This is because
of the condition of semiclassicity $S_0/\hbar \gg 1$ starts to fail
when diminishing $\beta_L$. This, in its turn, is caused by
decreasing of the barrier height $U_0$ and therefore the action
$S_0$. The analysis of the dependences $S_0(\beta_L)/\hbar$ reveals
that $S_0/\hbar \sim 1$ at $\beta\sim \beta_{L0}\;$ and the relative
divergence between the numerical and the analytical results for ScS
and 2JJ qubits is within 2 to 10 percent. For a SIS qubit a fit of the numerical %
and analytical results requires the more accurate fulfilment of the
semiclassicity condition. However, it is follows even from this
analysis that obtaining the tunnel splitting $\Delta E_{01}\gtrsim
1\,$K in the flux qubit based on a single SIS junction is impossible
under condition of weak ($U_0-E_1\gg k_BT$) influence of thermal
fluctuations on the metastable states decay. The dependences
$S_0(\beta_L)/\hbar$ for 2JJ, ScS and SIS qubits are close to linear
ones, whose slope (the action $S_0$ from $\beta_L$ rate of increase)
being higher in the indicated order. The value of tunnel splitting
in the region of its exponential smallness
$S_0(\beta_L)/\hbar\gg 1$ diminishes in the same sequence. %

The points in the numerical curves corresponding to the equal
heights of the potential barriers ($U_0=9.64\,$K) are indicated by
arrows in Fig.\;\ref{fig04}(a). The corresponding values of the
parameter couples $(\beta_L,E_{01}(\beta_L)/k_B)$ for 2JJ
($\lambda=0.9$), ScS and SIS flux qubits are:
$(1.06,3.45\,\mbox{K})$, $(0.88,1.79\,\mbox{K})$, $(1.60,0.16\,\mbox{K})$. %
It is seen that, under this condition, the tunnel splitting in a 2JJ
qubit is about twice the splitting in a ScS qubit and more than 20
times higher than that of a SIS qubit. The curve for the tunnel
splitting in 2JJ lies completely above the curves for ScS and SIS
qubits, and the tunnel splitting for a 2JJ qubit reaches the value
of $3.45\,$K at $\beta_{L}\simeq 1>\beta_{L0}$. The advantages of a
ScS qubit if compared to a SIS qubit was thoroughly analyzed in Ref.
\onlinecite{Shnyrkov}. Note that yet more increase of the tunnel
splitting in a 2JJ qubit in comparison with a ScS qubit with the
matched parameters mentioned above results from the fact that their
potentials (at $\lambda \simeq 1$) practically coincide while the
effective mass $M$ in 2JJ qubit is less by a factor of about two.
Fig.\;\ref{fig04}(b) shows the dependence $\Delta
E_{01}(\beta_L)/k_B$ for the 2JJ qubit at several $\lambda$, and
also a "level line", the line $\Delta E_{01}(\beta_L)/k_B$
corresponding to equal height (9.64 K) of the potential barriers in
2JJ qubit with varying $\lambda$. The curve $\Delta
E_{01}(\beta_L)/k_B$ shifts right when decreasing $\lambda$, and the
smaller being the value of $\lambda$, the higher the tunnel
splitting at a fixed $\beta_L$. This, however, is due to the
lowering of the barrier height $U_0$ when decreasing $\lambda$ that
leads to the exponential rise of the thermal decay rate. Note that
when desymmetrizing the junctions a fit between the numerical and
the analytical curves gets worse because $S_0(\beta_L)/\hbar$
decreases. As seen from the plot, the value of the tunnel splitting
gradually diminishes while moving along the level line with the
equal height of the potential barriers towards the lower values of
the junction symmetry parameter $\lambda$ (and the higher
$\beta_L$).

\section{Conclusions}
It should be emphasized that the principal requirements to 2JJ flux
qubits, namely: $\lambda\simeq 0.9$; $C\lesssim 50\,$fF/$\mu$m$^2$;
$j_c\sim 10^3$ A/cm$^2$; $I_c\sim\;1$$\mu$A at the JJ area $S_J\sim
0.1\;$$\mu$m$^2$; $L\sim 0.3\,$nH, $\beta_L\sim 1$ can be met with
the present-day technology based on Nb, NbN, MoRe materials with
superconductivity gap $\Delta(0)\simeq 10\,$K (see, e.g., Ref.
\onlinecite{Zangerle}). One can notice in conclusion that 2JJ flux
qubit with large amplitude of tunnel splitting potentially has some
strong advantages: (i) weak sensitivity to the motion of charge in
traps; (ii) extremely fast excitation (pumping frequency) in
qubit-based readout as well as in computer circuits due to
considerable increasing of the quantum tunneling rate $\nu\sim
\Delta E_{01}$; (iii) macroscopically large energy relaxation times
$\tau_\varepsilon$ (see, e.g., Ref. \onlinecite{Shnyrkov1} and Refs.
therein); (iv) further improvement of qubit coherence
characteristics \cite{Smirnov}.

\acknowledgments %
We thank W. Krech, D. Born, A.V. Ustinov and S.V. Kuplevakhsky for
helpful discussions. The work was supported by the Grant
No.\,M/189-2007 within the "Nanophysics and nanoelectronics" program
of the Ministry of Education and Science of Ukraine.

\end{document}